\message{ 
>>>>>>  >>>>>>>  >>>>>>
Please, reply  b  to the next question}
\input harvmac.tex


\def\tilde{\widetilde}
\def\bar{\overline}
\def\*{\star}
\def\({\left(}			
\def\){\right)}		
\def\[{\left[}		
\def\]{\right]}

%
%
\def\frac#1#2{{#1 \over #2}}

\def\d{\partial}

\def\ket#1{ | #1 \rangle}
\def\bra#1{ \langle #1 |}

\def\2pi{\hbox{$2\pi i$}}

\def\dsl{\raise.15ex\hbox{/}\kern-.57em\partial}
\def\Dsl{\,\raise.15ex\hbox{/}\mkern-.13.5mu D}
%
%
		\def\CC{{\cal C}}

		\def\CR{{\cal R}}
		
\def\CV{{\cal V}}		
		
%

%


\def\IR{\relax{\rm I\kern-.18em R}}
\font\cmss=cmss10 \font\cmsss=cmss10 at 7pt
\def\IZ{\relax\ifmmode\mathchoice
{\hbox{\cmss Z\kern-.4em Z}}{\hbox{\cmss Z\kern-.4em Z}}
{\lower.9pt\hbox{\cmsss Z\kern-.4em Z}}
{\lower1.2pt\hbox{\cmsss Z\kern-.4em Z}}\else{\cmss Z\kern-.4em Z}\fi}
\def\inbar{\,\vrule height1.5ex width.4pt depth0pt}
\def\IB{\relax{\rm I\kern-.18em B}}
\def\IC{\relax\hbox{$\inbar\kern-.3em{\rm C}$}}
\def\ID{\relax{\rm I\kern-.18em D}}
\def\IE{\relax{\rm I\kern-.18em E}}
\def\IF{\relax{\rm I\kern-.18em F}}
\def\IG{\relax\hbox{$\inbar\kern-.3em{\rm G}$}}
\def\IH{\relax{\rm I\kern-.18em H}}
\def\II{\relax{\rm I\kern-.18em I}}
\def\IK{\relax{\rm I\kern-.18em K}}
\def\IL{\relax{\rm I\kern-.18em L}}
\def\IM{\relax{\rm I\kern-.18em M}}
\def\IN{\relax{\rm I\kern-.18em N}}
\def\IO{\relax\hbox{$\inbar\kern-.3em{\rm O}$}}
\def\IP{\relax{\rm I\kern-.18em P}}
\def\IQ{\relax\hbox{$\inbar\kern-.3em{\rm Q}$}}
\def\IGa{\relax\hbox{${\rm I}\kern-.18em\Gamma$}}
\def\IPi{\relax\hbox{${\rm I}\kern-.18em\Pi$}}
\def\ITh{\relax\hbox{$\inbar\kern-.3em\Theta$}}
\def\IOm{\relax\hbox{$\inbar\kern-3.00pt\Omega$}}


\def\d{{\rm d}}

\def\oh{{1\over 2}}\def\un{{\bf 1}}

\def\Ga{\alpha}

\def\Gl{\lambda}

\def\Gr{\rho}


\def\mod{{\rm mod\,}}
\def\d{{\rm d}}

\def\bra{\langle}\def\ket{\rangle}



\lref\DFLZ{P. Di Francesco, F. Lesage and J.-B. Zuber, hep-th/9306018
to appear in {\it Nucl. Phys.} } 

\lref\W{N. Warner, {\it $N=2$ Supersymmetric Integrable Models and
Topological Field Theories}, to appear in the proceedings of the 1992 
Trieste Summer School, hep-th/9301088. }

\lref\Arn{V.I. Arnold,  and S.M. Gusein-Zade, A.N. Varchenko, 
{\it Singularities of differentiable maps}, Birk\"auser, Basel 1985.}

\lref\MVW{E. Martinec, {\it Phys. Lett.} {\bf B217} (1989) 431; 
{\it Criticality, catastrophes and compactifications}, in 
{\it Physics and mathematics of strings}, V.G. Knizhnik memorial volume, 
L. Brink, D. Friedan and A.M. Polyakov eds., World Scientific 1990\semi 
C. Vafa and  N.P. Warner, {\it Phys. Lett.} {\bf B218} (1989) 51.}

\lref\LVW{W. Lerche, C. Vafa, N.P. Warner, {\it Nucl. Phys.} 
{\bf B324} (1989) 427.}

\lref\EY{T. Eguchi and S.-K. Yang, {\it Mod. Phys. Lett.} 
{\bf A5} (1990) 1693 .}

\lref\DVV{R. Dijkgraaf, E. Verlinde and H. Verlinde, {\it Nucl. Phys.} 
{\bf B352} (1991) 59.}

\lref\DFZ{P. Di Francesco and J.-B. Zuber, 
in {\it Recent Developments in Conformal Field Theories}, Trieste
Conference, 1989, S. Randjbar-Daemi, E. Sezgin and J.-B. Zuber eds., 
World Scientific 1990
\semi P. Di Francesco, {\it Int.J.Mod.Phys.} {\bf A7} (1992) 407.}

\lref\Verl{E. Verlinde, {\it Nucl. Phys.} {\bf B300} [FS22] (1988) 360. }

\lref\CV{S. Cecotti and C. Vafa, 
{\it On Classification of $N=2$ Supersymmetric Theories}, 
preprint HUTP-92/A064,SISSA-203/92/EP, hep-th/9211097.  }

\lref\MaW{P. Mathieu and M.A. Walton, {\it Phys. Lett.} 
{\bf B 254} (1991) 106.} 

\lref\FLMW{P. Fendley, W. Lerche, S.D. Mathur and N.P. Warner,
{\it Nucl. Phys.} {\bf B348} (1991) 66 .}

\lref\FMVW{P. Fendley, S.D. Mathur, C. Vafa and N.P. Warner, 
{\it Phys. Lett.} {\bf 243B} (1991) 257. }

\lref\LWIG{W. Lerche and N.P. Warner, 
in {\it Strings \& Symmetries, 1991}, N. Berkovits, H. Itoyama et al. eds, 
World Scientific 1992;
{\it Nucl. Phys.} {\bf B358} (1991) 571\semi
K. Intriligator, {\it Mod. Phys. Lett.} {\bf A6} (1991) 3543\semi
D. Gepner, {\it Foundations of Rational Quantum Field Theory, I}
preprint CALT-68-1825, hep-th/9211100. }

\lref\HS{H. Saleur, {\it Nucl. Phys.} {\bf B382} (1992) 486, 532.  }

\lref\NW{D. Nemeschansky and N.P. Warner, {\it Nucl. Phys.} {\bf B380} (1992) 
241.}

\lref\CVtat{S. Cecotti and C. Vafa, 
{\it Nucl. Phys.} {\bf B367} (1991) 359.}

\lref\FI{P. Fendley and K. Intriligator, {\it Nucl. Phys.} {\bf B372} 
(1992) 533 (1992); {\it ibid.} {\bf B380} (1992) 265.}

%
\font\tenbf=cmbx10

\font\ninerm=cmr9

\font\eightrm=cmr8
\font\eightit=cmti8
\font\germ=eufm10
\def\sectiontitle#1\par{\vskip0pt plus.1\vsize\penalty-250
 \vskip0pt plus-.1\vsize\bigskip\vskip\parskip
 \message{#1}\leftline{\tenbf#1}\nobreak\vglue 5pt}
\hsize=6.0truein
\vsize=8.5truein
\parindent=15pt
\nopagenumbers
\baselineskip=10pt

\line{\eightrm
Interface between Physics and Mathematics
\hfil}

\vglue 5pc
\baselineskip=13pt
\headline{\ifnum\pageno=1\hfil\else%
{\ifodd\pageno\rightheadline \else \leftheadline\fi}\fi}
\def\rightheadline{\hfil\eightit
N=2 Superconformal Theories and their Integrable Deformations
\quad\eightrm\folio}
\def\leftheadline{\eightrm\folio\quad
\eightit
J.-B. Zuber
\hfil}
\voffset=2\baselineskip
\centerline{\tenbf
$N=2$\hskip 0.1cm SUPERCONFORMAL \hskip 0.1cm THEORIES\hskip 0.1cm
 }
\centerline{\tenbf
AND \hskip 0.1cm THEIR \hskip 0.1cm INTEGRABLE\hskip 0.1cm DEFORMATIONS }
\vglue 24pt
\centerline{\eightrm
JEAN-BERNARD ZUBER
}
\baselineskip=12pt
\centerline{\eightit
Service de Physique Th\'eorique de Saclay,
}
\baselineskip=10pt
\centerline{\eightit
(Direction des Sciences de la Mati\`ere du Commissariat \`a l'Energie Atomique)}
\baselineskip=12pt
\centerline{\eightit
F-91191 Gif-sur-Yvette,  France }
\vglue 20pt
\centerline{  }
\centerline{\eightrm ABSTRACT}
{\rightskip=2.5pc
\leftskip=2.5pc
\eightrm\parindent=1pc \baselineskip=12pt
After a short review of the algebraic setting of ${\scriptstyle N=2}$
superconformal field theories, their chiral ring and their perturbations, 
I present some recent results on curious relations between the 
integrability of their perturbations 
and algebraic properties of their deformed chiral ring. 
\vglue12pt}
\baselineskip=13pt
\overfullrule=0pt
\font\germ=eufm10
\def\gt g{\hbox{\germ g}}

\vskip 0.5cm

%

\def\d{{\rm d}}
\def\oh{{1\over 2}}\def\un{{\bf 1}}
\def\LG{Landau-Ginsburg\ }
\def\bra{\langle}\def\ket{\rangle}
\def\td{t_{\textstyle .}}\def\xd{x_{\textstyle .}}
\def\ie{{\it i.e.\ }}\def\nind{\noindent}
\def\Ref#1{${}^{#1}$}
%


\newsec{Introduction}
\noindent 
Two-dimensional field theories with a $N=2$ supersymmetry appear to be a
paradigm of all the beauties of the current mathematical physics. On the
one hand, $N=2$ superconformal theories describe the
ground states of ``realistic'' string theories and they are tightly connected 
with important mathematical issues such as Calabi-Yau manifolds, mirror
symmetries, etc. On the other hand,  some of them may 
play a role in the statistical mechanics of polymers\Ref{\HS}.
The deformations of these theories and their integrability properties 
are other fascinating issues. Also, by a twisting procedure, these 
theories produce
topological (or ``cohomological'') field theories; coupled to gravity, 
the simplest of these
describe interesting problems of intersection forms on the moduli spaces 
of curves. As such they are related to matrix models of two-dimensional
quantum gravity, and to the realm of classical integrable hierarchies
of KP and KdV types\dots 

At any rate, their structure is so constrained that a wealth of exact 
results has been obtained, and it is likely that more beautiful patterns
are still awaiting to be discovered. It is the purpose of this note to exhibit
some recently found curious properties\Ref{\DFLZ}, that seem to point to 
relations between different features of these theories.

\newsec{A short review of $N=2$ superconformal theories}
\noindent 
The symmetry algebra of $N=2$ superconformal theories 
is generated by the energy-momentum tensor $T(z)$, a
$U(1)$ current $J(z)$ and {\it two} supersymmetry fermionic 
generators $G^{\pm}(z)$. These fields are expanded on their Laurent 
moments respectively $L_n$, $J_n$, $G^{\pm}_{r+\oh}$ ($n$ is integer, and so
is $r$ in the Neveu-Schwarz sector). The primary fields of a given theory 
are specified by the eigenvalues of $L_0$ and $J_0$, 
the conformal weight $h$ and the $U(1)$ charge $q$. There is a subset 
of primary fields endowed with peculiar properties, 
namely the {\it chiral} fields, annihilated by the generators 
$G^+_{-\oh}$, that satisfy  $h= \oh q$ and form a ring for the 
pointwise (non singular!) operator product expansion (see \Ref{\W} 
for a review and references)
\eqn\IIa{ \lim_{z'\to z}\phi_i(z) \phi_j(z') =  \(\phi_i\phi_j\)(z) .}

In a wide class of $N=2$ superconformal theories,  a description by a
Landau-Ginsburg superpotential $W$ is available. The latter is a function
of (some) chiral (super)fields $X_{\Ga}$ only, 
and has been argued \Ref{\MVW} to be a 
quasi-homogeneous polynomial of these fields, with an isolated critical
point at the origin, to be found in the lists 
of singularities (or ``catastrophes'')\Ref{\Arn}. 

Particularly striking is the case of the so-called ``minimal'' 
$N=2$ theories, that have a central charge $c<3$. There, the 
\LG potential is given by one of the $ADE$ singularities \Ref{\Arn}, 
\eqnn\IIaa
$$\eqalignno{A_n \qquad W&= {x^{n+1}\over n+1}\cr
	  D_{n+2}\qquad W&= {x^{n+1}\over 2(n+1)}+ xy^2\cr
	     E_6 \qquad W&= {x^3\over 3}+ {y^4\over 4}& \IIaa \cr
	     E_7 \qquad W&= {x^3\over 3}+ {xy^3\over 3}\cr
	     E_8 \qquad W&= {x^3\over 3}+ {y^5\over 5}\cr }$$
and this matches the $ADE$ classification of their modular invariant 
partition functions. Moreover, the chiral ring of such a theory
is isomorphic to the local ring of the singularity : the latter
describes the multiplication of polynomials modulo the gradients of
$W$. In the 
simplest one-variable case, where the chiral ring is generated by 
powers of a single field, let $W(x)={x^{n+1}\over n+1}$ 
denote the $A_n$-superpotential
as a polynomial of the indeterminate $x$ and then
$\CR \equiv {\IC[x]/ W'(x)}$ has a basis $\{1,x, \cdots, x^{n-1}\}$, 
in one-to-one correspondence with the chiral primary fields of
$U(1)$ charge $q={1\over n+1}\times \{0,1,\cdots, n-1\}$. 

A further fascinating property of the $N=2$ theories is their deep
connection with topological field theories (TFT's). By ``twisting''
the energy-momentum tensor $T$, {\it i.e.} by writing \Ref{\EY}
\eqn\IIb{T_{{\rm top}}= T(z)+\oh \partial J(z)}
the chiral fields are now assigned a vanishing weight $h-\oh q=0$ and
are indeed shown to have $z$-independent correlation functions
\eqn\IIc{\bra \phi_{i_1}\phi_{i_2}\cdots \phi_{i_n}\ket\ .}

It is important and interesting to study the deformations 
that respect the $N=2$ supersymmetry. This is provided 
by the perturbations by ``top components'' of superfields
\eqn\IId{\bra \phi_{i_1}\phi_{i_2}\cdots \phi_{i_n}
e^{ -\sum_l t_l \int \d ^2 z\, G_{-\oh}^-\bar G_{-\oh}^- \phi_l(z,\bar z) 
+{\rm h.c.
}}\ket\ .}
We shall focus on perturbations by relevant operators, \ie such that 
$h_{\phi_l}< \oh \Leftrightarrow 
h_{G^-_{-\oh}\phi_l} <1$. The resulting theories
are no longer critical, they develop mass scales and may possess solitons 
interpolating between vacua. The other objects defined previously, chiral
ring, \LG potential and TFT, are also deformed accordingly.
When it exists, the \LG potential is deformed into a polynomial of the
$X$'s with coefficients functions of the $t$ parameters, denoted 
$W(X_{\Ga}, t_i)$. 
This deformation of the original homogeneous potential is endowed 
with remarkable properties, and the parameters $\td$ that provide 
this deformation are called ``flat coordinates''. 
The polynomials $p_i(\xd, \td)=-{ \partial W \over  \partial t_i}$, 
with $p_0=1$, form a basis of the deformed chiral ring and satisfy
\eqn\IIe{p_i(\xd,\td) p_j(\xd, \td) = C_{ij}^{\ \ k}(\td) p_k(\xd,\td)
\qquad \mod\, \partial_{\xd} W(\xd,\td) \ .}
The structure constants $C_{ij}^{\ \ k}$, beside the properties of
commutativity and associativity, satisfy two kinds of constraints\par
\item{$\bullet$} the metric tensor defined as $\eta_{ij}=C_{ij}^{\ \ 0}$
is independent of the $t$'s; 
\item{$\bullet$} $C_{ij}^{ \ \ k}$ 
satisfy the integrability conditions that enable
one to write $C_{ijk}=C_{ij}^{\ \ l}\eta_{kl}= 
{\partial^3\over \partial t_i\partial t_j\partial t_k}
F(\td)$, where $F(\td)$ is some function, the free energy of the theory. 

\nind 
Both assertions 
may be proven as consequences of the Ward identities of the $N=2$ 
theory \Ref{\DVV}. The former expresses that the propagator of the TFT 
is $t$-independent, whereas the latter constrains the three-point
functions of the perturbed TFT, that may be used to reconstruct all the 
($z$-independent) $n$-point functions 
$\bra \phi_{i_1}\phi_{i_2}\cdots \phi_{i_n}\ket$
$$\eqalignno{ \bra \phi_{i}\phi_{j}\cdots \phi_{k}\ket&=C_{ijk}\cr
\bra \phi_{i}\phi_{j}\phi_k \phi_{l}\ket &=C_{ijm}C_{kl}^{\ \ m}=
C_{ikm}C_{jl}^{\ \ m}, \qquad{\rm etc}\ . \cr}$$

On the other hand, integrability of some of the perturbations 
has been discussed by various authors \Ref{\refs{\FMVW\FLMW\MaW\CVtat
\FI{-}\NW}} : 
in some well chosen cases, 
the spectrum  of masses and the scattering theory of the solitons  may be 
computed exactly.  The ground states of the theory are given by the minima
of the bosonic potential, related to the superpotential by 
$V= |W'|^2$, and correspond thus to the extrema of $W$. Solitons 
interpolate between these ground states, and, in integrable 
cases, their masses are expressed in terms of the variation of $W$.
Recently, it has been pointed out \Ref{\DFLZ} following some earlier work 
\Ref{\LWIG}  that there  exists a curious connection
between the integrability and some algebraic properties 
of the chiral ring. This will be the subject of the next section.


\newsec{The pair of dual algebras}
\subsec{The normalizability of the $C$ matrices and the dual algebra.}
\nind
There is a certain similarity between the perturbed 
chiral algebra and the fusion algebra of rational conformal field theories 
(RCFT) that expresses how 
the primary fields merge in the operator product expansion.
This fusion algebra is expressed through the Verlinde
formula \Ref{\Verl}  in terms of a unitary matrix $S$ 
that encodes the modular properties of the RCFT
\eqn\IIIa{ N_{ij}^{\ k}=\sum_l {S_{il}S_{jl}S_{kl}^*\over S_{0l}}\ .}
In this formula, the two 
indices of the matrix $S$ are on the same footing, labelling the primary
fields of the theory and ``0'' refers to the identity operator. 
A corollary of this expression is that the 
commuting matrices $N_i$ of matrix elements $N_{ij}^{\ \ k}$ are
diagonalized in the basis provided by the orthonormal vectors
$S_j^{\ l}$.

In contrast with this fusion algebra, nothing guarantees that the 
deformed chiral 
algebra $C_{ij}^{\ \ k}(\td)$ may be diagonalized, not to speak of the 
orthogonality of eigenvectors. Let us now {\it assume} that, after a 
possible diagonal ($U(1)$ charge preserving) change of basis, the
matrices $C_i$ (of elements $C_{ij}^{\ \ k}$) may be diagonalized in 
an orthonormal basis $\psi_i ^{\, (a)}$
\eqna\IIIb
$$\eqalignno{ 
C_{ij}^{\ \ k}&={\Gr_i\Gr_j\over \Gr_k}M_{ij}^{\ \ k}& \IIIb a \cr
\(M_i\)_j^{\ k} &=\sum_a \Gl_i ^{(a)} \psi^{(a)}_j\psi^{(a)*}_k & \IIIb b \cr}
$$
(Beware that the positions of indices have been interchanged with respect to
\Ref{\DFLZ}~!) Because of the symmetry $i \leftrightarrow j$, the 
condition $p_0=1$ hence $M_0=\un$  and the orthonormality
of the $\psi$'s, one finds that the eigenvalues have the form
$\Gl_i ^{(a)}={\psi_i^{(a)}\over \psi_0^{(a)}}$ hence
\eqn\IIIc{ \(M_i\)_j^{\ k} =\sum_a {\psi^{(a)}_i \psi^{(a)}_j\psi^{(a)*}_k
\over\psi^{(a)}_0}\ . }
For simplicity, let us restrict to a case where the \LG potential depends 
on a single variable $x$ ; extension to more variables involves only 
notational difficulties. 
If $C_1$ denotes the matrix that encodes the multiplication by $x$ in the 
ring $	\CR= {\IC[x]/ \partial_x W(x)}$, 
\eqn\IIId{x p_i(x)= C_{1i}^{\ \ j}p_j(x)\qquad \mod\, \partial_x W(x) }
one sees that an extremum $x^{(a)}$ of $W$, hence a zero of $\partial_x W$, 
is an eigenvalue of $C_1$, given, according to \IIIb{} by
\eqn\IIIe{ x^{(a)}= \rho_1 {\psi_1^{(a)}\over\psi_0^{(a)}}\ . }

The property for a matrix $M$ to be diagonalizable in an orthonormal basis
is equivalent to its commutativity with its adjoint $[M,M^{\dagger}]=0$. 
Such a matrix is called {\it normal}, and we shall call a matrix 
made normal by a diagonal change of basis a {\it normalizable} matrix. 

When this is true, previous experience with such algebras \Ref{\DFZ}
suggests to also consider
the {\it dual} algebra, also associative and commutative, 
with structure constants given by a 
formula similar to \IIIb{} but with the alternative summation
\eqn\IIIf
{ N_{ab}^{\ \ c}=\sum_{\ell=0}^{n-1}{\psi_{\ell}^{(a)} \psi_{\ell}^{(b)} 
\psi_{\ell}^{(c) \,*} \over \psi_{\ell}^{(0)} }\ .}
where we assume that there is an eigenvector $\psi^{(0)}$ of the $M$'s 
with non vanishing components. 
(In the case of fusion, because of the hermiticity of $S$ up to charge 
conjugation $S=S^{\dagger}\CC$, the two algebras coincide.)

\subsec{Two intriguing observations}

\noindent
The property of the matrices $C$ to be normalizable turns out to be very
restrictive. When only one parameter $t_l$ is non vanishing, one may prove 
easily that normalizability in the $A_n$ case 
is only possible for perturbations by $t_1$, $t_2$ or $t_{n-1}$. These
cases, however, are also three cases for which the perturbed
$N=2$ theory is
known to admit an infinite number of conserved quantities and thus to 
be integrable\Ref{\refs{\LVW,\FLMW\CV{-}\FI}}, 
and according to a discussion of the semi-classical limit\Ref{\MaW}, they
are presumably the only ones. Thus one sees the 
amazing coincidence between two {\it a priori} quite different properties:
integrability of the perturbation of the $N=2$ theory 
and normalizability of the $C$ matrices in the TFT. 
This observation has prompted us \Ref{\DFLZ} 
to examine the perturbations of the 
other minimal $ADE$ models and of some others, for which the perturbed chiral
ring is known. In all cases (with a single non vanishing $t$), the same
agreement has been found! Normalizability was found for all cases that had 
been identified as integrable and only for those.
At the time of writing, this empirical observation remains fairly mysterious.
Could it be that these $N=2$ theories admit a specially simple criterion of
integrability? Notice that this property is independent of the existence 
of a \LG potential. 

Secondly, for all these normalizable cases, the two dual algebras have
been computed explicitly. A new curious feature appears: while the
structure constants of the $M$ algebra are in general non integers, the
$N$'s are integers! 
In some cases they are actually all non negative integers, 
in others they may be of either sign, but then, there is at
least one matrix $N_a$, $a\ne 0$, with non negative entries. This $N_a$
may be regarded as the adjacency matrix of a graph, and, new surprise!, 
this graph resembles the pattern of extrema of the 
potential. For example, for all the $ADE$ cases of \IIaa, the least 
relevant perturbation is normalizable ($t_{n-1}\ne 0$ in the $A_n$ case), 
and then one of the $N$'s turns out to be the adjacency matrix of the
corresponding $ADE$ Dynkin diagram, whereas the extrema of the 
perturbed potential form a pattern with the same topology. 
That there is a connection between the extrema and the eigenvectors 
of the Cartan matrices was the original observation of Lerche and Warner 
\Ref{\LWIG} that instigated the present work. This generalizes to all other 
normalizable perturbations. This is
apparent on Table I which summarizes all our findings on 
the normalizable perturbations of the $ADE$ potentials.

\bigskip
\bigskip
\bigskip

\nind {\bf Table I: $A, D, E$ perturbations.}

{\ninerm\baselineskip=12pt
\noindent{}Normalizable perturbations of the $A$, $D$, $E$
potentials are displayed as follows. First column: name
of the potential; second column: name of the perturbation given by 
the non--vanishing $t$ labelled by the $U(1)$ charge (times the 
Coxeter number) of the perturbing field; in the $D_{n+2}$ case, $\tau$ 
refers to the extra perturbation of charge  $n/2(n+1)$; 
third column: the dual ring
$N_{ab}^{\ \ c}$, through the graph of one of its generators;
fourth column: locus of the extrema of the perturbed potential
($x$ in the complex plane in the one variable case, $x$ and $y$
planes otherwise); fifth column: values taken by the perturbed potential
at the various extrema, in the complex plane (the links correspond 
to the minimal solitons interpolating between the extrema);
sixth column: a check--mark in case of known integrability, a question--mark
otherwise. }



\input epsf
\vfill\eject
{\par\begingroup\parindent=0pt\leftskip=1cm
\rightskip=1cm\parindent=0pt
\baselineskip=11pt
\epsfxsize=12cm
\midinsert
\centerline{\epsfbox{f1.eps}}
\bigskip
\centerline{\bf Table I.}
\par
\endinsert
\endgroup
\par}

\vfill\eject
{\par\begingroup\parindent=0pt\leftskip=1cm
\rightskip=1cm\parindent=0pt
\baselineskip=11pt
\epsfxsize=12cm
\midinsert
\centerline{\epsfbox{f2.eps}}
\bigskip
\centerline{\bf Table I (continued).}
\par
\endinsert
\endgroup
\par}

\vfill\eject
{\par\begingroup\parindent=0pt\leftskip=1cm
\rightskip=1cm\parindent=0pt
\baselineskip=11pt
\epsfxsize=12cm
\midinsert
\centerline{\epsfbox{f3.eps}}
\bigskip
\centerline{\bf Table I (continued).}
\par
\endinsert
\endgroup
\par}

\newsec{Discussion}

\noindent
It might seem that there is some resemblance of the present considerations
with the recent discussion of Cecotti and Vafa \Ref{\CV}. The latter 
authors have based a program of classification of $N=2$ theories 
on the analysis of the ``Dynkin diagram'' that encodes the 
intersection of the vanishing cycles of the (resolved) singularity \Ref{\Arn}.
Such a Dynkin diagram of a singularity 
is always connected and is intrinsically attached to the singularity 
(up to changes of the homology basis). In contrast,
our graphs are non necessarily connected. Moreover, they are defined 
only for certain (``normalizable'') perturbations and do depend on 
them; it seems that this dependence cannot be accounted for by a
simple change of basis. Thus one is led to conclude that the
two kinds of graphs are not related in an immediate way. 

What is the meaning of 
the integral $N_{ab}^{\ \ c}$? 

A curious observation in this respect is a similarity with the algebras
of characters and classes in a finite group. Let $G$ be a finite group 
of order $n$. Let $\chi^{(a)}_i$ denote the value of the character of 
the irreducible representation $(a)$ for the class $\CC_i$; finally let 
$n_i$ be the number of elements of $\CC_i$. The characters are known
to satisfy the orthogonality conditions
%
$ \sum_i {n_i\over n}\chi_i^{(a)}\chi_i^{(b)\,*}= \delta_{ab}$ 
from which we infer that the matrix $\psi_i^{(a)}=\sqrt{{n_i\over n}}
\chi_i^{(a)}$ is unitary. Then
%
\eqn\IVb{N_{ab}^{\ \ c}= \sum_i {\psi_i^{(a)}\psi_i^{(b)}\psi_i^{(c)\, *}
\over \psi_i^{(0)} }}
is the integer multiplicity of occurrence of the representation $(c)$ in 
the tensor product of $(a)$ and $(b)$, whereas the multiplicity 
$\tilde M_{ij}^{\ \ j}$ in the decomposition of the product of classes
$\CC_i \times \CC_j$ into $\CC_k$ is given by the dual algebra $M$, {\it up to the
rescaling by the $\sqrt{n}$'s  }
\eqn\IVc
{\tilde M_{ij}^{\ \ k}=\sqrt{n_i n_j\over n_k} M_{ij}^{\ \ k}
=\sqrt{n_i n_j\over n_k} \sum_a {\psi_i^{(a)}\psi_j^{(a)}\psi_k^{(a)\, *}
\over \psi_0^{(a)} }\ . }
So, in a finite group, 
although the $M$'s are not integers, they are made integers by a
rescaling by square roots of integers. It is an amusing fact that in all
the 
normalizable perturbations of $N=2$ theories considered so far 
\Ref{\DFLZ}, the same is true. In particular, it is
a non trivial property, (or so it seems to me), that the ``$M$-algebra''
of the $D$ and $E$ Dynkin diagrams can be made integral 
by such a change of basis. This seems to suggest that we are in the 
presence of some deformation of a finite group structure. The implications 
of this structure remain to be seen.


\newsec{Acknowledgements.}
\vbox{\noindent The work reported here has been carried out 
in a very friendly and enjoyable collaboration 
with Ph. Di Francesco and F. Lesage. 
I am also quite thankful to N. Warner for introducing me to
$N=2$ theories. }
\penalty -5000

\bigskip




\footatend\immediate\closeout\rfile\writestoppt
\baselineskip=12pt{\noindent{\bf 6. References}}\medskip{\frenchspacing%
\parindent=20pt\escapechar=` \input refs.tmp\vfill\eject}\nonfrenchspacing

\medbreak

%
\end